\documentclass[a4paper,10pt,notitlepage]{article}

\usepackage{amsfonts,amssymb,amsmath,amsthm,hyperref,colonequals,mathtools,xspace}
\usepackage[mathscr]{euscript}
\usepackage{lmodern}
\usepackage{cite}
\usepackage[T1]{fontenc}
\usepackage{bbm,bm} 
\usepackage{pdfsync}
\usepackage{xargs,xstring}
\usepackage[nice]{nicefrac} 
\usepackage{booktabs,multirow}

\theoremstyle{definition}
\newcommand{\beqa}{\begin{eqnarray}}
\newcommand{\eeqa}{\end{eqnarray}}
\newcommand{\beq}{\begin{equation}}
\newcommand{\eeq}{\end{equation}}
\newcommand{\g}{\ensuremath{\mathfrak{g}}\xspace}

\newcommand{\gl}[2]{\ensuremath{\mathfrak{gl}\left({#1}|{#2}\right)}}

\newcommand{\calH}{\mathcal{H}}

\newcommand{\calO}{\mathcal{O}}

\newcommand{\calN}{\mathcal{N}}

\newcommand{\calQ}{\mathcal{Q}}

\newcommand{\fg}{\mathfrak{g}}\newcommand{\fh}{\mathfrak{h}}

\newcommand{\fs}{\mathfrak{s}}

\newcommand\str{\text{str}}

\newcommand{\onehalf}{\ensuremath{\frac{1}{2}}\xspace}
\newcommand{\threehalves}{\ensuremath{\frac{3}{2}}\xspace}
\newcommand{\onethird}{\ensuremath{\frac{1}{3}}\xspace}

\newcommand{\Ya}{\ensuremath{\mathfrak{Q}}}
\newcommand{\Yas}{\ensuremath{\mathfrak{Q}^{\dagger}}}
\newcommand{\ya}{\ensuremath{\mathfrak{q}}}
\newcommand{\yas}{\ensuremath{\mathfrak{q}^{\dagger}}}
\newcommand\gr{\text{G}}

\newcommand\Rmat{\text{R}}
\newcommand\para{\chi}
\newcommand\shift{\mathbb{S}}
\newcommand\dyn{a}
\newcommand\tra{\text{T}}
\newcommand\mon{\text{M}}
\newcommand\Ham{\text{H}}
\newcommand\ham{\text{h}}
\newcommand\proj{\text{P}}

\newcommand{\per}{\ensuremath{\mathbb{P}}\xspace}

\DeclareMathOperator{\Z}{\mathbb{Z}\xspace}

\newcommand{\ket}[1]{\ensuremath{\left|\, #1\right>}}

\newcommand{\form}[2]{\ensuremath{\left( #1, #2\right)}}

\newcommand{\sln}[1]{\ensuremath{\mathfrak{sl}( #1 )}\xspace}
\newcommand{\slmn}[2]{\ensuremath{\mathfrak{sl}( #1|#2)}\xspace}

\newcommand{\gln}[1]{\ensuremath{\mathfrak{gl}( #1 )}\xspace}

\newcommand{\slnm}[2]{\ensuremath{\mathfrak{sl}( #1|#2)}\xspace}
\newcommand{\glnm}[2]{\ensuremath{\mathfrak{gl}( #1|#2)}\xspace}
\newcommand{\ospnm}[2]{\ensuremath{\mathfrak{osp}( #1|#2)}\xspace}
\newcommand{\psunm}[2]{\ensuremath{\mathfrak{psu}( #1|#2)}\xspace}

\newcommand{\uqglnma}[2]{\ensuremath{U_q(\widehat{\mathfrak{gl}}( #1|#2))}\xspace}
\newcommand{\uqslna}[1]{\ensuremath{U_q(\widehat{\mathfrak{sl}}( #1 ))}\xspace}

\newcommand{\yglnm}[2]{\ensuremath{\mathcal{Y}(\mathfrak{gl}( #1|#2))}\xspace}
\newcommand{\fso}{\fs(1)}
\newcommand{\fsoi}{\fso^{-1}}
\newcommand{\C}{\ensuremath{\mathbb{C}}\xspace}
\newcommand{\dg}[1]{ {\lvert #1 \rvert} }
\newcommand*\phantomas[3][c]{%
\ifmmode 
\makebox[\widthof{$#2$}][#1]{$#3$}%
\else 
\makebox[\widthof{#2}][#1]{#3}%
\fi 
}
\newcommand{\hlbox}[1]{\underline{\ensuremath{#1} }}

\makeatletter
\g@addto@macro\bfseries{\boldmath}
\makeatother

\begin{document}

\thispagestyle{empty}
\setcounter{page}{0}
\begin{flushright}\footnotesize
\texttt{HU-Mathematik-13-26}\\
\texttt{HU-EP-13/74}\\
\vspace{0.5cm}
\end{flushright}
\setcounter{footnote}{0}

\begin{center}
{\Large\textbf{
Dynamic Lattice Supersymmetry\\ in $\gl{n}{m}$ Spin Chains 
}\par}
\vspace{15mm}

{\sc  David Meidinger,
Vladimir Mitev}\\[5mm]

{\it  Institut f\"ur Mathematik und Institut f\"ur Physik,\\ Humboldt-Universit\"at zu Berlin\\
IRIS Haus, Zum Gro{\ss}en Windkanal 6,  12489 Berlin, Germany
}\\[5mm]

\texttt{david.meidinger@physik.hu-berlin.de}\\
\texttt{mitev@math.hu-berlin.de}\\[25mm]

\textbf{Abstract}\\[2mm]
\end{center}
Supersymmetry operators that change a spin chain's length have appeared in numerous contexts, ranging from the AdS/CFT correspondence to statistical mechanics models.  In this article, we present, via an analysis of the Bethe equations, all homogeneous, rational and trigonometric, integrable $\glnm{n}{m}$ spin chains for which length-changing supersymmetry can be present. Furthermore, we write down the supercharges explicitly for the simplest new models, namely the $\slmn{n}{1}$ spin chains with the $(n-1)$-fold antisymmetric tensor product of the fundamental representation at each site and check their compatibility with integrability.


\newpage
\setcounter{page}{1}


\tableofcontents
\addtolength{\baselineskip}{3pt}

\section{Introduction}\label{sec:intro}

Since the seminal work of \cite{Minahan:2002ve}, integrable spin chains have become increasingly important in the field of gauge theory and, in particular, have been instrumental for the proposed solution of the spectral problem of $\calN=4$ super Yang-Mills, see \cite{Beisert:2010jr} and reference therein. Nevertheless, considerable conceptual and technical difficulties remain and the subject is far from closed. The spin chains arising from the perturbative treatment of supersymmetric gauge theories have properties that seem quite unusual from the point of view of the ``classical'' models like the Heisenberg chain, such as long-range interactions that, starting at higher loop order, wrap around the spin chain, as well as length-changing or ``dynamic'' symmetry generators. It is this latter property that will be the focal point of this article.

So far, the issue of length-changing symmetry generators has been approached in essentially two ways. First, one can attempt to construct the symmetry algebra perturbatively as in \cite{Beisert:2003ys,Beisert:2004ry,Zwiebel:2008gr}. The second option involves considering the chain in the infinite length limit. In the setup of the coordinate Bethe Ansatz, the length-changing operators then simply acquire a momentum dependence and act with a non-trivial braiding which allows for an exact solution for the relevant S-matrix, \cite{Beisert:2005tm,Beisert:2006qh,Gomez:2006va, Plefka:2006ze, Arutyunov:2006yd}. This asymptotic solution \cite{Beisert:2005fw} can then be used as an input of the thermodynamic Bethe Ansatz which provides a way to compute the finite size spectrum, see for instance \cite{Gromov:2009bc,Gromov:2009zb,Arutyunov:2010gb}. Unfortunately, this method does not provide an insight into the way the length-changing symmetry is realized at finite length. 

Length-changing symmetries are not unique to gauge theories and their related spin chains. Similar structures have also been discovered in statistical mechanics and condensed matter theory. Investigating lattice fermion models with exclusion rules that are constructed to possess supersymmetry~\cite{Fendley:2002sg,Fendley:2003je,Yang2004}, it was found that some of these models map to integrable spin chains for which the supersymmetry generators become length-changing. These results were later generalized to elliptic and higher spin models \cite{Hagendorf:2011ps,Hagendorf:2012fz}, in a fashion that made some of the underlying general structure more transparent. A closer look shows that these models are similar to the ones arising in some subsectors\footnote{Examples include the non-compact fermionic $\sln{2}$ \cite{Beisert:2004ry}, the $\psunm{1,1}{2}$ subsectors of $\calN=4$ SYM \cite{Beisert:2007sk} and the \ospnm{4}{2} subsector of ABJM \cite{Zwiebel:2009vb}.} of the gauge theories at one-loop, where the manifest spin chain symmetry algebra $\g$ commutes with the $\glnm{1}{1}$ algebra formed out of the Hamiltonian $\Ham$, the two length-changing supercharges $\Ya$, $\Yas$ and the length-measuring operator. A question thus presents itself: what are all the integrable models that posses such a structure? The reasons for investigating this problem are twofold. First, since all supercommutators of the charges close into $\g\oplus\glnm{1}{1}$, these models have an exact symmetry algebra and not one that is only known perturbatively up to some order in a coupling constant. Thus, they constitute nice toy models for an analysis of length-changing symmetries in spin chains. Second, the relation between integrability and the length-changing symmetry remains obscure. They can occur independently from each other and it is unclear what deeper meaning their simultaneous appearance in a given model has. A comprehensive analysis of all integrable models with length-changing supersymmetry could shed some light on the connection. To find promising candidates, we analyze the Bethe equations carefully and provide a set of necessary conditions for the presence of length-changing supersymmetry such that its algebra has the form $\g\oplus\glnm{1}{1}$. 

The present article is structured as follows. We first introduce the reader in section \ref{sec:dynsusy} to the basic elements of length-changing or ``dynamic'' supersymmetry in spin chains and review the main elements of \cite{Hagendorf:2011ps,Hagendorf:2012fz}, adapted to graded representations. In section \ref{sec:bethe}, we analyze the Bethe equations for the rational and trigonometric $\glnm{n}{m}$ spin chains and present the cases in which we can expect dynamic supersymmetry to be present. The simplest novel candidates, involving integrable $\slmn{n}{1}$ spin chains, are then considered separately and the corresponding supercharges written explicitly  in section \ref{sec:sln1}. Finally, we devote section \ref{sec:conclusion} to discussions and conclusions.

\section{Dynamic supersymmetry}
\label{sec:dynsusy}

In this section, we want to present the main elements of dynamic supersymmetry in spin chains.  Mainly, we shall review here the results of \cite{Hagendorf:2011ps,Hagendorf:2012fz}, adapted to the language of superspaces. The reader is referred to the original publications for proofs. We start with a super vector space $V$ and define the spin chain Hilbert space as
\beq
\label{eq:Hilbertspace}
\calH\colonequals \bigoplus_{n=1}^{\infty}\calH_L, \qquad \calH_L\colonequals V\otimes V\otimes \cdots \otimes V\equiv V^{\otimes L}.
\eeq
We have the shift operator $\shift_L:\calH_L\rightarrow \calH_L$ acting in the usual way on homogeneous elements
\beq
\shift_Lv_1\otimes \cdots\otimes v_L=(-1)^{|v_L|\sum_{i=1}^{L-1}|v_i|}v_L\otimes v_1\otimes \cdots \otimes v_{L-1},
\eeq
where $|v|=0$ for even and $1$ for odd elements. We shall make use of the grading generator $\gr$ whose action on homogeneous elements is $\gr v=(-1)^{|v|}v$. Note that we shall consistently use the usual tensor product and not the graded one and take care of the signs by inserting the grading operator $\gr$ explicitly. 

Generalizing \cite{Hagendorf:2012fz}, we allow the length-changing symmetry operator $\Ya=\sum_{L=1}^{\infty}\Ya_L$ with $\Ya_L:\calH_L\rightarrow\calH_{L+1}$, to be either even or odd with respect to the grading of $V$. To treat both cases jointly, we set 
\beq
\omega\colonequals (-1)^{|\Ya|+1},
\eeq
and use it to define the projection operators $\Pi_L$ as 
\beq
\label{eq:projectoromega}
\Pi_L\colonequals \frac{1}{L}\sum_{j=0}^{L-1}\omega^{j(L+1)}\shift_L^j.
\eeq
The length-changing operator acts non-trivially only on the subsector $\bigoplus_{L=1}^{\infty}\Pi_L \calH_L$ of the Hilbert space, made out of the subspaces that have shift eigenvalues $\omega^{L+1}$. Thus, if $\Ya$ is bosonic, it acts on the subspaces that are alternate-cyclic, i.e. with shift eigenvalue $(-1)^{L+1}$, while if it is fermionic it acts non-trivially only on the cyclic ones. While changing the grading of $V$ sends one case to the other, we shall consider both here for completeness. 

The full length-changing operator $\Ya$ is a sum of local ones. Given such a local length-changing operator $\ya:V\rightarrow V\otimes V$ with $|\ya|=|\Ya|$ that satisfies the nilpotency condition
\beq
\label{eq:localnilpotency}
\Bigl(\bigl(\ya\otimes \mathbbm{1}\bigr)+\omega \bigl(\gr^{|\ya|}\otimes \ya\bigr)\Bigr) \ya v=\chi\otimes v-v\otimes \chi, \qquad \forall v\in V
\eeq
for some fixed $\chi\in V\otimes V$, we define the operators $\ya_i$ as follows:
\beq
\ya_{i+1}=\omega \shift_{L+1}\ya_i\shift_L^{-1}, \qquad \ya_{0}\colonequals \omega \shift_{L+1}^{-1}\ya_1\shift_L=\omega^L \shift_{L+1}\ya_L,
\eeq
where $\ya_1$ is just $\ya\otimes \mathbbm{1}\otimes \cdots \otimes \mathbbm{1}$.  The global length-changing supersymmetry generator $\Ya_L:\calH_L\rightarrow \calH_{L+1}$ can then be expressed in three equivalent ways
\beq
\label{eq:definitionYa}
\Ya_L\colonequals \sqrt{\frac{L+1}{L}}\Pi_{L+1}\sum_{i=1}^L\ya_i=\sqrt{\frac{L}{L+1}}\sum_{i=0}^L\ya_i\Pi_{L}=\sqrt{L(L+1)}\Pi_{L+1}\ya_k\Pi_L,
\eeq
where $k\in\{1,\ldots,L\}$ is arbitrary. Thanks to the local condition \eqref{eq:localnilpotency}, one can show that $\Ya_L$ is ``nilpotent'' in the sense that 
\beq
\label{eq:globalnilpotency}
\Ya_{L+1}\Ya_L=0.
\eeq 
In most cases of interest, the space $V$ is a representation of some superalgebra $\fg$. One can show via a straightforward computation that $\Ya_L$ commutes with the action of $\fg$ on $\calH$, i.e. that
\beq
\Delta^{(L)}(X)\Ya_L-(-1)^{|\Ya||X|}\Ya_L\Delta^{(L-1)}(X)=0, \qquad \forall X\in \g
\eeq
if $\ya$ commutes with $\g$ up to a gauge transformation that vanishes on the restricted Hilbert space, i.e. we must have
\beq
\label{eq:localinvariance}
\big(\ya X -(-1)^{|X||\ya|}\Delta(X)\ya\big)v=m_X \otimes v-\omega (-1)^{|v||m_X|}v\otimes m_X, \qquad \forall v\in V,
\eeq
where $m_X$ is a $X$-dependent element of $V$ with $|m_X|=|X|+|\ya|$ and we have used the standard superalgebra coproduct $\Delta(X)= \Delta^{(1)}(X)=X\otimes \mathbbm{1}+\gr^{|X|}\otimes X$. Thus, the right hand side of \eqref{eq:localinvariance} is graded symmetric for $\ya$ even and graded antisymmetric for $\ya$ odd.

In order to obtain the Hamiltonian, we also need a length-lowering operator. In all cases that we shall consider here, this operator is given by the superadjoint of $\Ya$. This means that we introduce a scalar product on $V$, extend it to $V\otimes V$ and define the superadjoint of a general operator $\calO:V\rightarrow V\otimes V$ as
\beq
\label{eq:definitinoadjoint}
\form{\calO^{\dagger} u}{v}_V\colonequals (-1)^{|\calO||u|}\form{u}{\calO v}_{V\otimes V}, \qquad \forall v\in V, \ u\in V\otimes V.
\eeq
In this way, we obtain $\yas: V\otimes V\rightarrow V $ and by conjugating \eqref{eq:definitionYa} an expression for $\Ya^{\dagger}_L$, which is also nilpotent, $\Ya^{\dagger}_{L-1}\Ya^{\dagger}_L=0$. We can then define a Hamiltonian density as
\beq
\label{eq:localhamiltoniandensity}
\ham(\ya)=\omega \big(\yas\otimes \mathbbm{1}\big)(\gr^{|\ya|}\otimes \ya)+\omega (\gr^{|\ya|}\otimes \yas)\big(\ya\otimes \mathbbm{1}\big)+\ya\yas+\frac{1}{2}\big(\yas\ya\otimes \mathbbm{1}+\mathbbm{1}\otimes \yas\ya\big)
\eeq
which gives us a nearest neighbor Hamiltonian  for the whole chain
\beq
\label{eq:definitionHamiltonian}
\Ham_L=\sum_{i=1}^L\ham(\ya)_{i,i+1}.
\eeq
For our analysis, the hermiticity of the scalar product is not required and the associated Hamiltonians need not be hermitian. Many interesting models, such as the $\uqslna{3}$ spin chain with the fundamental representation at each site, better known as the Perk-Schultz model \cite{Perk:1981nb}, have non-hermitian Hamiltonians.  

In the sectors of appropriate cyclicity, this Hamiltonian  can be expressed as
\beq
\label{eq:HasQQdagger}
\Pi_L\Ham_L\Pi_L=\Ya_L^{\dagger}\Ya_L+\Ya_{L-1}\Ya_{L-1}^{\dagger},
\eeq
which is the reason behind the existence of degeneracies at different lengths, since it implies that
\beq
\label{eq:commutationsrelationHQ}
\Ham_{L+1}\Ya_L=\Ya_L\Ham_L, \qquad \Ham_{L}\Ya_L^{\dagger}=\Ya_L^{\dagger}\Ham_{L+1}.
\eeq
If we denote by $\text{Pr}_L$ the projectors $\calH\rightarrow \calH_L$ and define
\beq
\mathfrak{H}\colonequals\sum_{L=1}^{\infty}\Ham_L\text{Pr}_L, \quad \Ya\colonequals\sum_{L=1}^{\infty}\Ya_L\text{Pr}_L, \qquad \mathfrak{L}\colonequals\sum_{L=1}^{\infty}L \text{Pr}_L, 
\eeq
then we see that span$\{\mathfrak{H},\Ya, \Ya^{\dagger},\mathfrak{L}\}$ forms a $\glnm{1}{1}$ algebra that commutes with the action of $\fg$ on $\calH$, so that the full symmetry of the model is $\fg\oplus\glnm{1}{1}$.

Of course, for an arbitrary local nilpotent charge $\ya$, the Hamiltonian defined in \eqref{eq:definitionHamiltonian} will not be integrable. In order to check the integrability of $\Ham_L$, we need to find an appropriate $\Rmat$ matrix, i.e. an operator $\Rmat(u):V\otimes V\rightarrow V\otimes V$ that reduces to the permutation operator $\per$ at some regularity point that we take to be $u=0$ and that obeys the Yang-Baxter equation:
\beq
\label{eq:YangBaxter}
\Rmat_{12}(u)\Rmat_{13}(u+v)\Rmat_{23}(v)=\Rmat_{23}(v)\Rmat_{13}(u+v)\Rmat_{12}(u)\qquad \forall u,v\in \mathbb{C}.
\eeq
Then, using an auxiliary space at site zero of the spin chain, one constructs a monodromy matrix and a transfer matrix as
\beq
\label{eq:monodromyandtransfer}
\mon_L(u)\colonequals \Rmat_{0L}(u)\cdots \Rmat_{01}(u), \qquad \tra_L(u)\colonequals\str_0\mon_L(u),
\eeq
where $\str_0$ is the supertrace over the auxiliary space.
The transfer matrix reduces to the shift operator at the regularity point $u=0$, i.e. $\tra_L(0)=\shift_L$.  The nearest neighbor Hamiltonian is then uniquely defined, up to some arbitrary complex length-depending constants\footnote{These constants can be absorbed in the normalization of $\Rmat$. } $\alpha_L$ and $\beta_L$, as the logarithmic derivative of the transfer matrix at $u=0$ i.e.
\beq
\label{eq:integrableHamiltonian}
\Ham_L=\alpha_L \frac{d}{du}\log \tra_L(u)\vert_{u=0}+\beta_L\mathbbm{1}.
\eeq
If an $\Rmat$ matrix exists such that \eqref{eq:integrableHamiltonian} holds, we can write the Hamiltonian density of \eqref{eq:definitionHamiltonian}, up to two arbitrary complex constants, as the logarithmic derivative of the $\Rmat$ matrix itself. Thus, in order to make sure that the Hamiltonian density of \eqref{eq:localhamiltoniandensity} leads to an integrable problem, we need to find an $\Rmat$ matrix satisfying the Yang-Baxter equation \eqref{eq:YangBaxter} such that its logarithmic derivative at the regularity point gives $\ham(\ya)$.

We consider here the reverse problem, i.e. we look for integrable Hamiltonians for which a charge $\ya$ exists such that \eqref{eq:HasQQdagger} holds. If such a charge exists for an integrable spin chains, then we expect that its effects should manifest themselves at the level of the Bethe equations in the form of degeneracies in the energy spectrum for different lengths. 
Such degeneracies have indeed been observed in the case of some models for which the dynamic supersymmetry was already known \cite{Beisert:2005fw,Hagendorf:2011ps}. 
Here we proceed in reverse:
we \emph{identify} models with a hidden dynamic supersymmetry by systematically looking for the respective degeneracies
in the Bethe equations, for which the supercharges can then hopefully be constructed.
The next section is devoted to classifying these models in the context of $\glnm{n}{m}$ integrable spin chains.

\section{Analysis of the Bethe equations}
\label{sec:bethe}

In this section, we shall classify all periodic and homogeneous rational and trigonometric $\glnm{n}{m}$ spin chains for which dynamic supersymmetry in the sense of section \ref{sec:dynsusy} is expected to be present by analyzing their Bethe equations. We begin by writing down the equations for arbitrary representations of $\glnm{n}{m}$ and analyze extensively some low rank examples that already contain all the subtleties of the problem. Subsection \ref{sec:classificationhigherrank} then presents the general result.

Before we begin, let us  introduce some notation related to the algebra $\glnm{n}{m}$.
The fundamental representation is spanned by the vectors $\{e_i\}_{i=1}^{n+m}$, $n$ of which are even while
$m$ are odd. We abbreviate the gradation by $|i|\colonequals |e_i|$. The generators $\{E_{ij}\}_{i,j=1}^{n+m}$
with degree $|E_{ij}|=|i|+|j|$ (mod 2) obey the super commutation relations
\beq
[E_{ij},E_{kl}] = \delta_{jk} E_{il} - (-1)^{(|i|+|j|)(|k|+|l|)} \delta_{il}E_{kj}.
\eeq
In the fundamental representation, the generators are given by $(n+m)\times(n+m)$ matrices with entries
$(E_{ij})_{kl}=\delta_{ik}\delta_{jl}$.
The representations $V$ that we are interested in are all highest weight and we denote by $\lambda_i$ the
eigenvalue of the Cartan generator $E_{ii}$ on the highest weight vector.
Finally, we introduce the normalized Dynkin labels
\beq
\label{eq:Dynkinlabel}
a_i\colonequals (-1)^{|i|}\lambda_i-(-1)^{|i+1|}\lambda_{i+1},
\eeq
which will be very useful for the subsequent discussion. Some common examples of representations include the fundamental representation with $\lambda_i=\delta_{i1}$ and $a_i=(-1)^{|1|}\delta_{i1}$ and the adjoint representation with $\lambda_i=\delta_{i1}-\delta_{i(n+m)}$ and  $a_i=(-1)^{|i|}\delta_{i1}+(-1)^{|i+1|}\delta_{i(n+m-1)}$. Thus, for $\mathfrak{gl}(2)$, $a_1=1$, respectively $2$, for the fundamental, respectively the adjoint, i.e. $a_1$ is twice the spin.


\subsection{Bethe equations for \texorpdfstring{$\glnm{n}{m}$ spin chains}{gl(n|m)}}

The initial ingredient is an $\Rmat$ matrix, which in our case has either a $\yglnm{n}{m}$ or a $\uqglnma{n}{m}$ symmetry, corresponding to rational or trigonometric $\glnm{n}{m}$ spin chains respectively.  This $\Rmat$ matrix differs from the one of section~\ref{sec:dynsusy}, since it intertwines between the fundamental representation of the corresponding quantum group and the representation $V$ from which the spin chain is built. One constructs the monodromy $\mon$ and transfer matrix $\tra$ as in \eqref{eq:monodromyandtransfer}, with the fundamental representation placed in the auxiliary space and diagonalizes the transfer matrix $\tra$ using the Nested Bethe Ansatz \cite{Belliard:2008di}. One can of course not obtain the Hamiltonian directly in this way, since the $\tra$ thus constructed does not reduce to the shift operator for any value of the spectral parameter, i.e. there is no regularity point. One can however construct the eigenvalues for the transfer matrices stemming from different auxiliary spaces from the fundamental one via fusion, so that degeneracies of the eigenvalues of $\tra$ imply degeneracies in the eigenvalues of the Hamiltonian.

The Bethe equations we will investigate can be found in \cite{Belliard:2008di,Frassek:2011aa,Ragoucy:2007kg,Arnaudon:2004vd,Arnaudon:2005sg} and in a compact notation read
\begin{equation}
        \frac{\Lambda_{k+1}(u_j^{(k)}\!+\frac{c_k}{2})}
        {\Lambda_k(u_j^{(k)}\!+\frac{c_k}{2})}
        =\gamma_k
        \frac{\calQ_{k-1}(u_j^{(k)}\!-\frac{(-1)^{|k|}}{2})}
        {\calQ_{k-1}(u_j^{(k)}\!+\frac{(-1)^{|k|}}{2})}
        \frac{\calQ_{k}(u_j^{(k)}\!+(-1)^{|k|})}
        {\calQ_{k}(u_j^{(k)}\!-(-1)^{|k+1|})}
        \frac{Q_{k+1}(u_j^{(k)}\!-\frac{(-1)^{|k+1|}}{2})}
        {\calQ_{k+1}(u_j^{(k)}\!+\frac{(-1)^{|k+1|}}{2})},
        \label{eq:betheequations}
\end{equation}
for $j=1,\cdots,M_k$ and $k=1,\cdots,n+m-1$. We remind that the superscript $(k)$ refers to the level of the nesting and the subscript $j$ labels the distinct root of a single level, running from $1$ to the total number $M_k$. We use the abbreviations
\beq
\gamma_k\colonequals (-1)^{|k|+|k+1|+1}, \qquad c_k\colonequals\sum_{i=1}^k (-1)^{|i|}
\eeq 
and choose our conventions such that the constant factors in the arguments of the $\calQ$ functions are real. A given set of Bethe roots obeying these equations uniquely determines an eigenvector of the transfer matrix with eigenvalue
\begin{equation}
        \Lambda(u)=\frac{1}{N^L}\sum_{k=1}^{n+m}(-1)^{|k|}\Lambda_k(u)
        \frac{\calQ_{k-1}(u-\frac{c_k}{2}-\frac{(-1)^{|k|}}{2})}
        {\calQ_{k-1}(u-\frac{c_k}{2}+\frac{(-1)^{|k|}}{2})}
        \frac{\calQ_{k}(u-\frac{c_k}{2}+(-1)^{|k|})}
        {\calQ_{k}(u-\frac{c_k}{2})},
        \label{eq:transfermatrixeigenvalues}
\end{equation}
where we have introduced a normalizing factor $N$. We will parametrize the $q$-deformation, depending on the notational convenience, by either $q$, $\eta$ or $\zeta$, which are related by $q=e^\eta$ and $\eta=\frac{2\pi i}{\zeta}$. Defining the function 
\beq
\fs(x)\colonequals \sinh\bigl(\eta x\bigr), 
\eeq
the $\calQ$-functions and the weights are given by
\begin{align}
        \calQ_i(u)&=\prod_{j=1}^{M_i}(u-u_j^{(i)}),&
\Lambda_i(u)&=\bigl(u-(-1)^{|i|}\lambda_i\bigr)^L & 
& \mbox{for }\;\yglnm{n}{m},\notag\\
\calQ_i(u)&=\prod_{j=1}^{M_i}\frac{\fs(u-u_j^{(i)})}{\fso},&
\Lambda_i(u)&=\Bigl( \frac{\fs(u-(-1)^{|i|}\lambda_i)}{\fso} \Bigr)^L &
& \mbox{for }\;\uqglnma{n}{m}.
        \label{eq:qsandweights}
\end{align}

\subsection{Low rank examples}\label{sec:lowrank}

Dynamic supersymmetry reveals itself by certain degeneracies in the Bethe equations.
Before we present the general case, it is useful to first familiarize oneself with
the general structure of these degeneracies
by considering some simple low rank case, namely $\gln{2}$, $\gln{3}$ and $\glnm{2}{1}$.

\subsubsection*{Models based on \texorpdfstring{\gln{2}}{gl(2)}}

The Bethe equations\footnote{Due to our normalizations, we can consider the trigonometric and rational case together, since no divergences appear for $\eta\rightarrow 0$.} \eqref{eq:betheequations} for $\gln{2}$ can be written as
\begin{equation}
        \left( \frac{\fs(u_j+\onehalf-\lambda_2)}
                    {\fs(u_j+\onehalf-\lambda_1)} \right)^L
        =-
        \prod_{i=1}^{M}
        \frac{\fs(u_j-u_i+1)}
             {\fs(u_j-u_i-1)},\quad j=1,\cdots,M.
        \label{eq:bethegltwo}
\end{equation}
Now imagine we have a set of $M$ Bethe roots satisfying these equations
for length $L$. We shall now try to obtain a solution for length $L-1$, using the same set of roots plus an additional one
$u_\ast$.  The resulting $M+1$ Bethe equations then read
\begin{equation}
        \left( \frac{\fs(u_j+\onehalf-\lambda_2)}
                    {\fs(u_j+\onehalf-\lambda_1)} \right)^L
               \frac{\fs(u_j+\onehalf-\lambda_1)}
                    {\fs(u_j+\onehalf-\lambda_2)}
        =-
        \prod_{i=1}^{M}
        \frac{\fs(u_j-u_i+1)}
             {\fs(u_j-u_i-1)}
        \frac{\fs(u_j-u_\ast+1)}
             {\fs(u_j-u_\ast-1)},
        \label{eq:constraintsgltwozero}
\end{equation}
for $j=1,\cdots,M$ with one additional equation for $u_\ast$
\begin{equation}
        \label{eq:constraintsgltwoone}
        \left( \frac{\fs(u_\ast+\onehalf-\lambda_2)}
        {\fs(u_\ast+\onehalf-\lambda_1)} \right)^{L-1}
        =
        \prod_{i=1}^{M}
        \frac{\fs(u_\ast-u_i+1)}
             {\fs(u_\ast-u_i-1)}.
\end{equation}
We focus on the first $M$ equations. Using our knowledge
that $u_1,\cdots,u_M$ satisfy the Bethe equations of length $L$,
we reduce them to
\begin{equation}
        \frac{\fs(u+\onehalf-\lambda_1)}
             {\fs(u+\onehalf-\lambda_2)}
        =
        \frac{\fs(u-u_\ast+1)}
             {\fs(u-u_\ast-1)}
        \;\Longrightarrow\,
        \left\{\begin{array}{l}        
        \fs(u+\tfrac{1}{2}-\lambda_1)=
        \fs(u-u_\ast+1)\\
        \fs(u+\tfrac{1}{2}-\lambda_2)=
        \fs(u-u_\ast-1)
        \end{array}\right. ,
        \label{eq:constraintsgltwothree}
\end{equation}
where the left hand side of \eqref{eq:constraintsgltwothree}  is equivalent to the set of two equations on the right hand side, since it has to be valid for arbitrary $u\in\C$.
This requirement stems from the demand that the individual Bethe roots be essentially arbitrary.
Using the period $\zeta$ of the function $\fs$, we have
\begin{equation}
        \onehalf-\lambda_1=-u_\ast+1+m_1\zeta,\qquad
        \onehalf-\lambda_2=-u_\ast-1+m_2\zeta,
        \label{eq:constraintsgltwofour}
\end{equation}
where here and in the following all $m_i$ are integers. Taking the sum and the difference of the last two equations
we arrive at
\begin{equation}
        m_3\zeta=\dyn_1+2,
        \qquad
        u_\ast=\onehalf(\lambda_1+\lambda_2-1+m_4\zeta),
        \label{eq:constraintsgltwofive}
\end{equation}
where the normalized Dynkin label was introduced in \eqref{eq:Dynkinlabel}.
Finally, the Bethe equation for the new root $u_\ast$ simply reduces to the condition
\begin{equation}
        \prod_{i=1}^{M}\frac{\fs(u_i+\tfrac{1}{2}-\lambda_2)}
                             {\fs(u_i+\tfrac{1}{2}-\lambda_1)}
                             =(-1)^{L-1}
        \label{eq:momentumconstaintgltwo}
\end{equation}
which forces the states to lie in the alternate-cyclic sectors, just as required for a bosonic length-changing operator, see \eqref{eq:projectoromega}.

We will now show that the two Bethe states share the same transfer matrix eigenvalue, if the transfer matrix
is properly normalized.
For the case at hand the eigenvalue of the transfer matrix \eqref{eq:transfermatrixeigenvalues} is
just
\begin{equation}
\label{eq:normalizedgl2transfereigenvalue}
        \Lambda(u)=
        \left(\frac{\fs(u-\lambda_1)}{\fso N}\right)^L
        \prod_{i=1}^M \frac{\fs(u-u_i+\onehalf)}
        {\fs(u-u_i-\onehalf)}
        +\left(\frac{\fs(u-\lambda_2)}{\fso N}\right)^L
        \prod_{i=1}^M \frac{\fs(u-u_i-\threehalves)}
        {\fs(u-u_i-\onehalf)}.
\end{equation}
Going to $L-1$ and adding the extra root we find the eigenvalue
\begin{align}
\label{eq:normalizationL}
        \Lambda_\ast(u)=&
        \left(\frac{\fs(u-\lambda_1)}{\fs(1)N}\right)^L
        \prod_{i=1}^M \frac{\fs(u-u_i+\onehalf)}
        {\fs(u-u_i-\onehalf)}
        \left[ 
                \frac{\fs(u-u_\ast+\onehalf)}{\fs(u-u_\ast-\onehalf)}
                \frac{\fs(1)N}{\fs(u-\lambda_1)}
        \right]\notag\\
        +&\left(\frac{\fs(u-\lambda_2)}{\fs(1)N}\right)^L
        \prod_{i=1}^M \frac{\fs(u-u_i-\threehalves)}
        {\fs(u-u_i-\onehalf)}
        \left[ 
                \frac{\fs(u-u_\ast-\threehalves)}{\fs(u-u_\ast-\onehalf)}
                \frac{\fs(1)N}{\fs(u-\lambda_2)}
        \right].
\end{align}
By \eqref{eq:constraintsgltwofour}, the normalization $N=\fsoi\fs(u-\lambda_2+1)$
makes the terms inside the square brackets of \eqref{eq:normalizationL} equal to $1$, so that $ \Lambda_\ast(u)= \Lambda(u)$.

To recapitulate, we see that if the Dynkin label and the $q$ deformation are related by $\dyn_1+2= m \zeta $, $m\in \Z$, then, given a solution to the Bethe equations at length $L$, we can construct a solution at length $L-1$ that shares the same eigenvalue of the transfer matrix by adding an extra root $u_\ast$. 

When applying this procedure, one needs to keep two limitations in mind. First, in some spin chains like the rational $\gln{2}$ one with $a_1=1$, two Bethe roots cannot be equal. In those cases, we cannot add $u_\ast$ if it is already contained in the set of Bethe roots, in which case the solution that we started with must have descended from a solution at length $L+1$ that does not contain $u_\ast$. Thus, one can expect that the symmetry is nilpotent and that the two states are in a doublet. In other spin chains, due to the possible presence of repeated\footnote{As noted in \cite{Avdeev:1985cx, Hao:2013rza}, physical solutions with repeated roots appear in rational $\gln{2}$ spin chains with $a_1\geq 2$. The treatment of such cases requires the introduction of a regulator and exceeds the scope of this work. For the trigonometric case, the reader is referred to  \cite{Baxter:2001sx}.} roots, we cannot prove that the symmetry implied by the degeneracy should be nilpotent. Since the $\Ya$ charges that we want to construct are nilpotent by definition, the appearance of degeneracies in the Bethe equations is a necessary but not a sufficient condition for the presence of a symmetry of the type presented in section \ref{sec:dynsusy}. In our experience however, even in spin chains for which repeated roots are a possibility, like the rational $\gln{2}$ spin chain with $a_1=2$, having a degeneracy of the energies at different lengths implies the presence of a nilpotent length-changing operator.  The second limitation prevents us from adding $u_\ast$ if it would bring us beyond the allowed number of roots, meaning that there are states that are singlets under the symmetry.

Furthermore, we need to be careful when solving the condition $\dyn_1+2= m \zeta $, because the points $q=\pm 1$ are singular for the algebra $\uqglnma{n}{m}$, even though the Bethe equations are well-defined. Since $q=e^{2\pi i/\zeta}$, we need to exclude $\zeta=1/m^{\prime}$ and $\zeta=2/m^{\prime}$, $m^{\prime}\in \Z\backslash\{0\}$. Thus, for the $q$-deformed spin chains, we exclude the Dynkin labels that obey $\dyn_1+2= \pm 1,\pm 2$. In the rational limit, the Bethe equations do not contain periodic functions, but the same conditions for the presence of supersymmetry hold as before with the restriction that all the integers $m_i$ are set to zero, i.e. the representations for which supersymmetry is present in the rational limit are also those for which it is present for any value of the $q$-deformation. Therefore, we find only one possible representation for which length-changing supersymmetry is present in the rational limit, namely the non-compact one with $a_1=-2$. The supercharges in this case were constructed in~\cite{Beisert:2004ry}.

Summarizing, for the trigonometric models, we find the following representations for which length-changing supersymmetry is expected to be present
\begin{itemize}
        \item $\dyn_1+2$ is a generic complex number, corresponding
                to an infinite dimensional representation 
                of \gln{2} with $q=e^{\frac{2 \pi i m}{a_1+2}}$, $m\in \Z\backslash\{0\}$. 
        \item $\dyn_1+2$ is an integer, which now implies that $q$ 
                has to be some root of unity satisfying $q^{a_1+2}=1$. This includes all finite dimensional representations, except the trivial one, since $q^2\neq 1$.  
\end{itemize}

So far, we have considered transformations decreasing the length while adding a root. A peculiarity of the rank one case is the existence of transformations sending $L\to L+1$.
Since the derivation of the corresponding constraint equations
is similar, we will just state the results:
The constraints are
\begin{equation}
        m_1\zeta=\dyn_1-2,
        \qquad
        u_\ast=\onehalf(\lambda_1+\lambda_2-1+m_2\zeta).
        \label{eq:constraintsgltwolplusone}
\end{equation}
Furthermore, the normalization of the transfer matrix \eqref{eq:normalizedgl2transfereigenvalue} which makes the two solutions of the Bethe equations have the same eigenvalue is
\begin{equation}
        N=\frac{\fs(u-\lambda_2-2)\;\fs(u-\lambda_2)}{\fso\fs(u-\lambda_2-1)}.
\end{equation}

In particular the $\dyn_1=2$, or spin 1, is the only finite dimensional $\gln{2}$
representation that exhibits dynamic supersymmetry
in the rational limit and its supercharges can be found in \cite{Hagendorf:2012fz}. We also find solutions in the 
$q$-deformed case starting from $a_1=5$ with $q^{a_1-2}=1$.
A summary of all these values can be found in table~\ref{tab:dynsusysl2}.

\begin{table}
\begin{tabular}{ c c c c c c c c c c c c }
        \toprule
        Spin $\nicefrac{a_1}{2}$ & & $0$ & \nicefrac{1}{2} & $1$ & \nicefrac{3}{2} & $2$ & \nicefrac{5}{2} & $3$ & \nicefrac{7}{2} & $\!\cdots$\,\\
        \midrule
        $L\to L-1$ & &  --- & $q^3=1$ & $q^4=1$ & $q^5=1$ & $q^6=1$ & $q^7=1$ & $q^8=1$ & $q^9=1$ & $\cdots$ \\[2pt]
        $L\to L+1$ & &  --- & --- & all $q$ & --- & --- & $q^3=1$ & $q^4=1$ & $q^5=1$ & $\!\cdots$\\
        \bottomrule
\end{tabular}
\caption{Values of the deformation parameter $q$ for which higher spin XXZ models exhibit dynamic supersymmetry.}
\label{tab:dynsusysl2}
\end{table}

\subsubsection*{Models based on \texorpdfstring{\gln{3}}{gl(3)}}

We now turn to the first algebra with rank greater than one,
for which the Bethe equations are nested. For \gln{3} they read
\begin{align}
        \left( \frac{\fs(u_j^{(1)}\!+\onehalf-\lambda_2)}
                    {\fs(u_j^{(1)}\!+\onehalf-\lambda_1)} \right)^L
&\!=-
\prod_{i=1}^{M_1} \frac{\fs(u_j^{(1)}\!-u_i^{(1)}\!+1)}
                        {\fs(u_j^{(1)}\!-u_i^{(1)}\!-1)}
\prod_{i=1}^{M_2} \frac{\fs(u_j^{(1)}\!-u_i^{(2)}\!-\onehalf)}
                        {\fs(u_j^{(1)}\!-u_i^{(2)}\!+\onehalf)}
,\quad j=1,\cdots,M_1\notag\\
\left( \frac{\fs(u_j^{(2)}\!+1-\lambda_3)}
            {\fs(u_j^{(2)}\!+1-\lambda_2)} \right)^L
&\!=-
\prod_{i=1}^{M_1} \frac{\fs(u_j^{(2)}\!-u_i^{(1)}\!-\onehalf)}
                        {\fs(u_j^{(2)}\!-u_i^{(1)}\!+\onehalf)}
\prod_{i=1}^{M_2} \frac{\fs(u_j^{(2)}\!-u_i^{(2)}\!+1)}
                        {\fs(u_j^{(2)}\!-u_i^{(2)}\!-1)}
,\quad j=1,\cdots,M_2
\end{align}
Our strategy remains the same, namely, we change the length and insert a new root $u_\ast$, which can now be of type $1$ or $2$. Let us denote the level at which the new root is inserted by $\ell$. As one can straightforwardly check, the equations one has to solve
to find dynamic supersymmetry with $L\to L+1$ have no solution,
so we will consider only $L\to L-1$ and discuss the two possibilities, $\ell=1$ and $\ell=2$, jointly. 

Along the lines of the discussion of \gln{2}, we have to solve the following system for the $\lambda$'s and $u_\ast$, for arbitrary
$u_j^{(k)}$'s:%
\begin{align}
        \phantomas[l]{\fs(u_j^{(2)}+\onehalf-\lambda_2)\,}
        {\hlbox{\ell=1}},
        &&
        \phantomas[l]{\fs(u_j^{(2)}+\onehalf-\lambda_2)\,}
        {\hlbox{\ell=2}}
        &\notag\\
        \fs(u_j^{(1)}+\onehalf-\lambda_1)
        &=\fs(u_j^{(1)}-u_\ast+1),&
        \fs(u_j^{(1)}+\onehalf-\lambda_1)
        &=\fs(u_j^{(1)}-u_\ast-\onehalf),\notag\\
        \fs(u_j^{(1)}+\onehalf-\lambda_2)
        &=\fs(u_j^{(1)}-u_\ast-1),&
        \fs(u_j^{(1)}+\onehalf-\lambda_2)
        &=\fs(u_j^{(1)}-u_\ast+\onehalf),\notag\\
        \fs(u_j^{(2)}+1-\lambda_2)
        &=\fs(u_j^{(2)}-u_\ast-\onehalf),&
        \fs(u_j^{(2)}+1-\lambda_2)
        &=\fs(u_j^{(2)}-u_\ast+1),\notag\\
        \fs(u_j^{(2)}+1-\lambda_3)
        &=\fs(u_j^{(2)}-u_\ast+\onehalf),&
        \fs(u_j^{(2)}+1-\lambda_3)
        &=\fs(u_j^{(2)}-u_\ast-1),
        \label{eq:constraintsglthreezero}
\end{align}
which can be rewritten in a form that separates the constraints on the weights from the required
value of the extra root,
\begin{align}
        {\hlbox{\ell=1}}
        &&
        \dyn_1+2&=m_1\zeta,&
        \dyn_2-1&=m_2\zeta,&
        u_\ast&=\onethird(\lambda_1+\lambda_2+\lambda_3-\threehalves+m_3\zeta),\notag\\
        {\hlbox{\ell=2}}&&
        \dyn_1-1&=m_1\zeta,&
        \dyn_2+2&=m_2\zeta,&
        u_\ast&=\onethird(\lambda_1+\lambda_2+\lambda_3-1+m_3\zeta).
        \label{eq:constraintsglthree}
\end{align}

At this point it is easy to see why the corresponding equations for $L\to L+1$ have no solutions. Note that in going from \eqref{eq:constraintsglthreezero} to \eqref{eq:constraintsglthree},
the number of equations has been reduced. This was caused by the equality of the two equations
involving $\lambda_2$. If we had chosen to send $L\to L+1$, this would not have been the case. Even worse,
the two equations would have been 
$\lambda_2-u_\ast=-\onehalf+m_1\zeta$ 
and 
$\lambda_2-u_\ast=\onehalf+m_1\zeta$, which together imply that  $q=1$, a value that we need to exclude.

As in the rank one case, the right hand side of the Bethe
equations for $u_\ast$ becomes the shift eigenvalue of the original state if \eqref{eq:constraintsglthreezero} is fulfilled. Since we are allowed to perform global shifts in the spectral parameter, the Bethe equations only depend on the Dynkin labels. This can be used to show that the shift eigenvalue of the original state has to be $(-1)^{L-1}$. Thus, the length-changing supersymmetry is still present only for the alternate-cyclic states, just as in the rank one case. Finally, the normalization of the transfer matrix that is necessary to make supersymmetry doublets have the same eigenvalue is
\beq
N=\Lambda_3(u) \text{ for } \ell=1, \qquad N= \Lambda_1(u)  \text{ for } \ell=2.
\eeq
We can now discuss the different solutions to \eqref{eq:constraintsglthree}.
\begin{itemize}
        \item If the Dynkin labels are
                generic complex numbers, there will
                be no value of $\zeta$ for which the first
                two equations of \eqref{eq:constraintsglthree} can hold simultaneously.
        \item If the $\lambda_i$'s are integers, we set $\zeta=\frac{A}{m}$ and recast \eqref{eq:constraintsglthree} into the form
\beq
        \hlbox{\ell=1}  \qquad
        \frac{A}{m}=
        \frac{\dyn_1+2}{m_5}=
        \frac{\dyn_2-1}{m_6}, \qquad 
        \hlbox{\ell=2}  \qquad
        \frac{A}{m}=
        \frac{\dyn_1-1}{m_5}=
        \frac{\dyn_2+2}{m_6}.
\eeq
               Here, $A$ and $m$ are in $\Z$. Stated in this form, one can see that the possible values of $\zeta$ are obtained if $A$ is set to be
the greatest common divisor of the two numerators:
\begin{align}
        \hlbox{\ell=1} & \qquad
        A\equiv\gcd(\dyn_1+2,\dyn_2-1)\neq 1, 2 \; , \notag\\
        \hlbox{\ell=2} & \qquad
        A\equiv\gcd(\dyn_1-1,\dyn_2+2)\neq 1, 2 \; ,
        \label{eq:restrictionsuqslthreeb}
\end{align}
where we imposed the constraint $q^2\neq 1$. Thus, the possible values of $q$ are given by the formula $q^A=1$. There are many representations satisfying these constraints. As an example, one can take the two series
                $\lambda_1\geq 2$, $\lambda_2=1$, and its dual
                $\lambda_1\geq 2$, $\lambda_2=\lambda_1-1$, 
                with $q$ given in each case by
                $q^{\lambda_1+1}=1$.
It should be stressed that the constraints have no solution for the fundamental representation.
        \item The rational solutions to \eqref{eq:constraintsglthree}, obtained by setting the
                $m_i$ to zero, are infinite dimensional
                and have \sln{3} Dynkin labels 
                $[\dyn_1,\dyn_2]=
[-2,1]$ for $\ell=1$ and $[1,-2]$ for $\ell=2$.
\end{itemize}

\subsubsection*{Models based on \texorpdfstring{\glnm{2}{1}}{gl(2|1)}}
It is well known that Lie superalgebras admit
different choices of Cartan matrices that lead
to different gradations. In our experience, considering different gradations does not lead to novel models, so we will restrict the discussion to the distinguished gradation of \glnm{2}{1} with even indices $1$ and $2$ and an odd index $3$. In this gradation, the Bethe equations read
\begin{align}
        \left( \frac{\fs(u_j^{(1)}\!+\onehalf-\lambda_2)}
                    {\fs(u_j^{(1)}\!+\onehalf-\lambda_1)} \right)^L
&\!=-
\prod_{i=1}^{M_1} \frac{\fs(u_j^{(1)}\!-u_i^{(1)}\!+1)}
                        {\fs(u_j^{(1)}\!-u_i^{(1)}\!-1)}
\prod_{i=1}^{M_2} \frac{\fs(u_j^{(1)}\!-u_i^{(2)}\!-\onehalf)}
                        {\fs(u_j^{(1)}\!-u_i^{(2)}\!+\onehalf)}
,\quad j=1,\cdots,M_1\notag\\
\left( \frac{\fs(u_j^{(2)}\!+1+\lambda_3)}
            {\fs(u_j^{(2)}\!+1-\lambda_2)} \right)^L
&\!=
\prod_{i=1}^{M_1} \frac{\fs(u_j^{(2)}\!-u_i^{(1)}\!-\onehalf)}
                        {\fs(u_j^{(2)}\!-u_i^{(1)}\!+\onehalf)}
,\quad j=1,\cdots,M_2.
\end{align}
Insertion of the extra root at the ``bosonic'' level ($\ell=1$)
works very much as before. Again there is only the possibility
to go from $L$ to $L-1$ and the constraint equations read
\begin{align}
        \dyn_1+2&= m_1\zeta, &
        \dyn_2-1&=m_2\zeta,   &
        u_\ast&=\onethird(\lambda_1+\lambda_2-\lambda_3-\threehalves+m_3\zeta).
\end{align}
In addition we find that the transfer matrix normalization is given by $N=\Lambda_3(u)$ and that the supersymmetry maps between states with shift eigenvalues $(-1)^{L-1}$.

If we want to insert the extra root at the ``fermionic'' level $\ell=2$,
there is a new subtlety. Due to the lack of self-scattering of
magnons of type 2, the left hand side of the Bethe equations for
this level has to become constant, meaning that it is not allowed to carry momentum.
This gives us a first constraint:
\begin{equation}
        \dyn_2=m_1\zeta.
        \label{eq:constrsl21a}
\end{equation}
The other constraints, arising from actual
cancellations between momentum and scattering factors
read
\beq
        \dyn_1\mp 1=m_2\zeta,\qquad
        u_\ast=\onehalf(\lambda_1+\lambda_2-1+m_3\zeta),
        \label{eq:constrsl21b}
\eeq
for $L\to L\mp 1$. Thus, there are two important differences compared to the cases discussed so far. First, because there is only one cancellation involving each $\lambda$, both directions of the change of length are possible. Second, the Bethe equation for the extra root tells us that the shift eigenvalue of the original state is one, i.e. we are restricted to the cyclic sectors. The necessary normalization of the transfer matrix is $N=\Lambda_1(u)$. As for \gln{2} and \gln{3} it is easy to list  the representations and values of the deformation parameter that satisfy the constraints \eqref{eq:constrsl21a} and \eqref{eq:constrsl21b}.
Restricting to representations with integer weights, one finds
\begin{itemize}
        \item two rational solutions with Dynkin labels $[\pm1,0]$ for $L\to L\mp 1$.  In section \ref{sec:sln1}, we shall present the supercharges for the model with the fundamental representation explicitly.
        \item $q$-deformed solutions that have to satisfy $A=\gcd(\dyn_1\mp 1,\dyn_2)\neq 1,2$ for $L\to L\mp1$. The deformation is then determined by $q^A=1$.
\end{itemize}

\subsection{Classification for arbitrary rank}
\label{sec:classificationhigherrank}

After having gained some experience by considering
low rank examples, we are now ready to tackle the
general case. 
The aim of this section is to extend the classification
of integrable models with dynamic supersymmetry 
to all rational and $q$-deformed models based on 
\glnm{n}{m} with arbitrary gradation and any representation.

The results obtained so far suggest a very nice
strategy to achieve this: we will start by considering the
slightly easier case of rational models, for which all 
cancellations have to hold directly, without exploiting 
any periodicities. This will yield a set of constraints 
$\{X_j(\lambda)=0\}$.
We can then immediately deduce the results for $q$-deformed
chains: In principle any representation works, but we have to
require\footnote{Note that in this scheme we are not allowed to rescale the constraints $X_j=0\to \mathrm{const}\times X_j=0$.} that 
\begin{align}
        A=\gcd(\{X_j\})\neq 1,2.
        \label{eq:qstrat1}
\end{align}
If this is satisfied, $q$ is forced to be an $A$'th root of unity, $q^A=1$.

Before discussing the actual equations that need to be solved
it is useful to examine some general requirements that have to
be met.

\subsubsection*{General remarks}

For higher rank algebras, the cancellations can only work if the level $\ell$, at which the extra Bethe root is inserted, scatters with all momentum carrying levels and only with those. If it does not self-scatter, the level $\ell$ is also not allowed to carry momentum.
For the Bethe equations under consideration, this implies that only the insertion level and its nearest neighboring ones are allowed to carry momentum.
Also, for \glnm{n}{m}, self-scattering is absent precisely if
the level corresponds to a fermionic node in the Dynkin diagram of \slnm{n}{m}, i.e. if $\dg{\ell}\neq\dg{\ell+1}$.
Moreover, the momentum factors on the left hand side of level $\ell$  vanish exactly when $\dyn_\ell=0$.

Thus, without doing any calculations, we can restrict the Dynkin
labels of representations with dynamic supersymmetry to be of the
form 
\beq
[\dyn_1,\ldots,\dyn_{\ell-1},\dyn_{\ell},\dyn_{\ell+1},\ldots,\dyn_{n+m-1}]
=\begin{cases}
        [0,\ldots,0,\#,\#,\#,0,\ldots,0] & \text{for } \dg{\ell}=\dg{\ell+1}\\
        [0,\ldots,0,\#,0,\#,0,\ldots,0] & \text{for } \dg{\ell}\neq\dg{\ell+1}
\end{cases}.
\label{eq:dynkinsusy}
\eeq
We can now proceed to determine the actual values of the non-zero Dynkin labels
separately for the two cases.

\subsubsection*{Insertion at a bosonic level}

To fix $\dyn_{\ell-1}$, $\dyn_{\ell}$ and $\dyn_{\ell+1}$,
we have to check all cancellations with the extra root, of which there
are six, see \eqref{eq:betheequations}. Just as in the $\gln{3}$, we have to discard the possibility to change the length $L\to L+1$. For $L\to L-1$, the constraints in the rational case read
\begin{equation}
       \dyn_{\ell-1} =(-1)^\dg{\ell},\qquad 
       \dyn_{\ell} =-\bigl((-1)^\dg{\ell}+(-1)^\dg{\ell+1}\bigr),\qquad 
        \dyn_{\ell+1}=(-1)^\dg{\ell+1},
      \label{eq:constraintsbosrat}
\end{equation}
and determine the value of the additional Bethe root 
\begin{align}
        u_\ast=\frac{1}{4}\left(
        \sum_{j=\ell-1}^{\ell+2}(-1)^{\dg{j}}\lambda_j
        -c_\ell-c_{\ell+1}
        \right).
        \label{eq:extrarootbosrat}
\end{align}

Now we can apply the argument given around equation \eqref{eq:qstrat1}.
If we $q$-deform the Bethe equations, any Dynkin labels work, as long as they satisfy
\beq
        A=\gcd\bigl[\dyn_1,\ldots, \dyn_{\ell-1}-(-1)^\dg{\ell}, \dyn_{\ell}
        +\bigl( (-1)^\dg{\ell}+(-1)^\dg{\ell+1}\bigr), \dyn_{\ell+1}
        -(-1)^\dg{\ell+1},  \ldots, \dyn_{n+m-1} \bigr]
        \neq 1,2.
\eeq
Then the valid deformation parameters are given by $q^A=1$ and the extra Bethe root is given by $u_\ast+\frac{m}{4}\zeta$, $m\in \Z$, where $u_\ast$ is the one of \eqref{eq:extrarootbosrat}.

Regarding the momentum sectors the supersymmetry requires,
it is an easy exercise to check that they are given by shift
eigenvalues $(-1)^{L-1}$, as in the \gln{2} and \gln{3} cases
discussed above, i.e. alternate-cyclic. It is also not difficult to show that the transfer matrix eigenvalues \eqref{eq:transfermatrixeigenvalues}
remain invariant under the addition of the extra root if they are normalized by $N=\Lambda_{\ell-1}(u)$. One merely has to apply the equalities in \eqref{eq:constraintsbosrat}, modulo periodicity in the $q$-deformed case, to the eigenvalue with the extra root.

\subsubsection*{Insertion at a fermionic level}
If the insertion level corresponds to a fermionic node in the Dynkin diagram we only need to determine $\dyn_{\ell\pm 1}$. Since $\dyn_{\ell}=0$, we see that there are no multiple cancellations involving $\lambda_\ell$ or $\lambda_{\ell+1}$ -- thus both $L\to L-1$ as well as $L\to L+1$ are possible. One easily determines the constraints
\beq
\dyn_{\ell-1}=\pm (-1)^\dg{\ell}, \qquad \dyn_{\ell+1}=\pm (-1)^\dg{\ell+1},
\eeq
for $L \to L \mp 1$;
the extra Bethe root is again given by \eqref{eq:extrarootbosrat} and is the same whether we send
$L\to L-1$ or $L\to L+1$.

For the $q$-deformed case, we obtain that for $L\to L\mp 1$, the weights need to satisfy the constraints
\beq
        A=\gcd\bigl[\dyn_1,
        \ldots, \dyn_{\ell-2},\dyn_{\ell-1}\mp(-1)^\dg{\ell},
        \dyn_{\ell}, \dyn_{\ell+1}\mp(-1)^\dg{\ell+1}, \dyn_{\ell+2},
        \ldots,
        \dyn_{n+m-1}\bigr]\neq 1,2
\eeq
with $q^A=1$, with the same $u_\ast$ as in the bosonic case. Furthermore, invariance of the transfer matrix eigenvalues is possible and requires a normalization $N=\Lambda_{\ell-1}(u)$.

\subsubsection*{Special cases}

For concreteness, we will now state the results for the most important cases, the bosonic algebras \gln{n} and for \glnm{n}{m} in the distinguished gradation explicitly.

For the bosonic Lie algebras \gln{n}, we can only decrease the 
length while inserting the extra root. The constraints
for insertion at level $\ell$ then
read
\beq
[\dyn_1,\ldots,\dyn_{\ell-2},\dyn_{\ell-1},\dyn_{\ell},\dyn_{\ell+1},\dyn_{\ell+2},\ldots,\dyn_{n-1}]
=
[0,\ldots,0,1,-2,1,0,\ldots,0].
\eeq
For the $q$-deformed case, simply use the formula \eqref{eq:qstrat1}. One should remember that the \gln{2} case is special and allows 
transformations $L\to L+1$, see section \ref{sec:lowrank}. Note in particular that all representations for which dynamic supersymmetry works
for the rational models are infinite dimensional, except the spin $1$ representation of \gln{2}, which is such
a special case with $L\to L+1$.

The distinguished gradation of \glnm{n}{m}
is 
\begin{equation}
        \dg{1}=\cdots=\dg{n}=0\qquad\text{and}\qquad\dg{n+1}=\cdots\dg{n+m}=1.
\end{equation}
The Dynkin diagram related to this choice of
grading has a single fermionic node at position $n$.
Thus all bosonic insertion levels are covered by the discussion
of \gln{n} above\footnote{If one keeps in mind to replace
$\lambda_j$ for $j\geq n+1$ by $-\lambda_j$ -- but this only affects
the constraint for level $n$.}
and we just have to look at the special insertion $\ell=n$,
for which the constraints read
\beq
[\dyn_1,\ldots,\dyn_{\ell-2},\dyn_{\ell-1},\dyn_{\ell},\dyn_{\ell+1},\dyn_{\ell+2},\ldots,\dyn_{n+m-1}]
=
[0,\ldots,0,\mp 1,0,\pm 1,0,\ldots,0]\; \text{for } L \to L \pm 1.
\eeq
Again one can apply \eqref{eq:qstrat1} to obtain representations for which the supersymmetry is present in 
$q$-deformed models.

In our conventions, see \cite{Frappat:1996pb}, a representation of \slnm{n}{m} is 
finite dimensional if all $\dyn_j$ are integers, except for $\dyn_n$ which 
can be any complex number, and if $\dyn_j\geq 0$ for $j=1,\ldots,n-1$ and
$\dyn_j\leq 0$ for $j=n+1,\ldots,n+m-1$.
Thus we find that representations which feature dynamic supersymmetry
in their rational models with $L\to L+1$ are infinite dimensional.
On the other hand, the $L\to L-1$ cases are all finite dimensional.
The nicest examples are the algebras \slnm{n}{1}, for which the fermionic node 
is at the boundary of the Dynkin diagram. The representation for these Dynkin labels corresponds to a Young diagram consisting 
of $n-1$ antisymmetrized boxes. In particular, the \slnm{2}{1} spin chain features dynamic supersymmetry for the fundamental representation, as we have already seen in \eqref{sec:lowrank}. In the next section we will construct the supercharges for these models.
\medskip

As a final remark, we observe that, if we add the extra root at level $\ell$ while changing the length as $L\rightarrow L\pm 1$, the representations for which dynamic supersymmetry is expected to be present in the rational limit are exactly those whose highest weight is equal to $\pm\alpha_{\ell}$, where the $\alpha_i$ are the simple roots. Remember that increasing the length is not always possible. This observation may shed some light on the generalization of this procedure to other algebras.

\section{The length-changing operators for \texorpdfstring{$\slmn{n}{1}$}{sl(n|1)}}
\label{sec:sln1}

In this section, we wish to construct explicitly the length-changing operators for one class of models, namely for the rational $\slmn{n}{1}$ spin chains having the $(n-1)$-fold antisymmetric tensor product of the fundamental representation at each site. We begin by reviewing the $\slmn{2}{1}$ case for which the solution was already given in \cite{Hagendorf:2012fz} and which we generalize to the $q$-deformed case. From this special case, it is then straightforward to construct the local length-changing operators for all $\slmn{n}{1}$ cases and compare the resulting Hamiltonian density with the one derived from integrability.

The methods presented in this section, namely obtaining the $\Rmat$ matrices by fusion and comparing the resulting Hamiltonian density with the one obtained from the $\ya$ supercharges derived by requiring certain invariance conditions, can be expected to be generalizable to all $\glnm{n}{m}$ spin chain examples found in section \ref{sec:bethe}. We hope to return to this general problem in a future work.

\subsection{The lowest rank example}
\label{subsec:tJ}

The simplest case involves the rational spin chain constructed using the fundamental representation of $\slmn{2}{1}$ at each site. This spin chain maps to the t-J model at specific values of the parameters, see for instance \cite{Fendley:2003je},  and in \cite{Hagendorf:2012fz} the local length-changing operator was given as
\beq
\label{eq:sl21supercharge}
\ya e_i =\delta_{i3}\big(e_2\otimes e_1-e_1\otimes e_2),
\eeq
where $e_i$ are the basis vectors of the fundamental representation, $e_3$ being fermionic.
The only non-vanishing entries of the superadjoint are then $\yas e_2\otimes e_1=-\yas e_1\otimes e_2=e_3$. Plugging this in the formula for the Hamiltonian density \eqref{eq:localhamiltoniandensity} immediately leads to 
\beq
\ham(\ya)=\mathbbm{1}-\per, 
\eeq
which can be written as $\mathbbm{1}-\para \partial_u\log(\Rmat(u))\vert{_u=0}$ using the normalized $\Rmat$ matrix $\Rmat(u)=\frac{1}{\para}(u+\para \per)$. In the above $\para$ is an arbitrary complex number that we keep for convenience. Thus we see that dynamic supersymmetry is present in this integrable spin chain.

Before moving on, we remark that the supercharge \eqref{eq:sl21supercharge} can be easily generalized to the $\mathfrak{U}_q(\slmn{2}{1})$ spin chain:
\beq
\ya e_i =\delta_{i3}\big(e_2\otimes e_1- q e_1\otimes e_2),
\eeq
for arbitrary non-vanishing $q$. This supercharge yields via \eqref{eq:localhamiltoniandensity} the correct $\mathfrak{U}_q(\slmn{2}{1})$ Hamiltonian density.

\subsection{Representation theory}
\label{subsec:representation}

We now wish to summarize the main aspect of the representation theory of $\slmn{n}{1}$ that we are going to make use of in the following.

Let $V_F$ be the fundamental representation of $\fg\colonequals \slmn{n}{1}$. Let $(e_i)_{i=1}^{n+1}$ be a basis of $V_F$, where the $\{e_i\}_{i=1}^n$ are bosonic and $e_{n+1}$ is fermionic. The set of covariant $\fg$ representations can be classified by Young diagrams that lie in the ``fat hook''. We shall denote a Young diagram that has $k_1$ rows of length $l_1$, $k_2$ rows of length $l_2$ and so on, with $l_i>l_{i+1}$, by $l_1^{k_1}l_2^{k_2}\cdots$. We shall consider in particular the $n-1$ antisymmetric tensor product of the fundamental representation given by the Young diagram $1^{n-1}$ and we shall call this representation $V$ in accordance with section \ref{sec:dynsusy}.

In order to give explicit expressions in this representation, we need a good basis. The standard one is given by 
\beq
e_{i_1}\wedge \cdots \wedge e_{i_{n-1}}\colonequals \frac{1}{r!}\sum_{\sigma\in S_{n-1}}(-1)^{|\sigma|}\per_{\sigma}\left(e_{i_1}\otimes \cdots \otimes e_{i_{n-1}}\right).
\eeq
We recall that the permutation $\per_{\sigma}$ will introduce additional signs when acting on fermions. For example, if $n=4$, then $e_5$ is fermionic and $e_5\wedge e_5\wedge e_5=e_5\otimes e_5\otimes e_5$. It is very useful to introduce another basis, that in a sense is obtained via a particle-hole transformation, as follows
\beq
\label{eq:definitionnewbasis}
\ket{k_1,\ldots, k_r}\colonequals \underbrace{e_1\wedge\cdots \wedge \hat{e}_{k_1}\wedge \cdots \wedge  \hat{e}_{k_2}\wedge \cdots  \wedge  \hat{e}_{k_r}\wedge \cdots \wedge e_n}_{n-r \text{ vectors}}\wedge \underbrace{e_{n+1}\wedge e_{n+1}\wedge \cdots \wedge e_{n+1}}_{r-1 \text{ vectors}},
\eeq
where $1\leq k_1<k_2\cdots<k_r\leq n$ and $\hat{\cdot}$ indicates omission. It is obvious that there are exactly 
$\binom{n}{r}$ elements of the type $\ket{k_1,\ldots, k_r}$ and that $r=1,\ldots, n$, which tells us that the dimension of $V$ is equal to $2^n-1$. It is furthermore clear that $\ket{k_1,\ldots, k_r}$ has the fermionic degree $r-1 \mod 2$ . 
The simple root raising and lowering operators act as
\beq
\begin{split}
\label{eq:actionofsymmetryholebasis}
E_{k_i,k_i+1}\ket{k_1,\ldots, k_r}&=\big(1-\delta_{k_i+1,k_{i+1}}\big)\ket{k_1,\ldots,\hat{k}_i,k_{i}+1\ldots,k_r},\\
E_{n,n+1}\ket{k_1,\ldots, k_r}&=\delta_{k_r,n}(r-1)\ket{k_1,\ldots,k_{r-1}},\\
E_{k_i,k_i-1}\ket{k_1,\ldots, k_r}&=\big(1-\delta_{k_i-1,k_{i-1}}\big) \ket{k_1,\ldots,k_i-1,\hat{k}_i,\ldots,k_r},\\
E_{n+1,n}\ket{k_1,\ldots, k_r}&=\big(1-\delta_{k_r,n}\big)\ket{k_1,\ldots,k_{r},n}.
\end{split}
\eeq
In order to write the $\Rmat$ matrix and the Hamiltonian density, we shall need the tensor product decomposition
\beq
\label{eq:tensordecompositionVV}
V\otimes V = 1^{n-1}\otimes 1^{n-1}\cong \bigoplus_{i=0}^{n-1}2^{n-1-i}1^{2i}.
\eeq
We now want to find the corresponding highest weight states.  Using the basis of \eqref{eq:definitionnewbasis}, we claim that the highest weight vectors $v_i$ for the representation $2^{n-1-i}1^{2i}$ are given by $v_0\colonequals \ket{n}\otimes \ket{n}$ and by
\begin{multline}
\label{eq:highestweightvectors}
v_i\colonequals\sum_{m=1}^i(-1)^{i(m-1)}\binom{i-1}{m-1}\sum_{\sigma\in S_{m,i+1-m}}(-1)^{|\sigma|}  \ket{n-i-1+\sigma(1),\ldots,n-i-1+\sigma(m)}\\\otimes \ket{n-i-1+\sigma(m+1),\ldots,n-i-1+\sigma(i+1)},
\end{multline}
for $i=1,\ldots, n-1$. In the above, we have made use of the following group quotient
\beq
\label{eq:defquotientS}
S_{i,j}\colonequals S_{i+j}/(S_i\times S_{j})
\eeq
in order to have the permutations keep the proper ordering. Note that in \eqref{eq:highestweightvectors} $\sigma(j)$ is taken to be an element of $\{1,\ldots,i+1\}$ for all $j$. 

We finish this subsection by considering the quadratic Casimir. We recall that given a basis $T_a$ of a Lie superalgebra with a scalar product $\kappa_{ab}=(T_a,T_b)$, the quadratic Casimir $\mathfrak{C}_2$ is given by $\mathfrak{C}_2=\kappa^{ab}T_aT_b$. A straightforward computation leads to the following general formula for the value of the properly normalized $\g$ quadratic Casimir in a general representation of highest weight $\lambda$:
\begin{multline}
\mathfrak{C}_2=-\frac{1}{2}\Big[\sum_{k=1}^n\big(\lambda_k-\frac{1}{n-1}\sum_{l=1}^{n+1}\lambda_l\big)\big(\lambda_k+n-2k-\frac{1}{n-1}\sum_{l=1}^{n+1}\lambda_l\big)\\-\big(\lambda_{n+1}+\frac{1}{n-1}\sum_{l=1}^{n+1}\lambda_l\big)\big(\lambda_{n+1}+n+\frac{1}{n-1}\sum_{l=1}^{n+1}\lambda_l\big)\Big].
\end{multline}
Thus, in the representation $2^{n-1-i}1^{2i}$, the quadratic Casimir gives the value $i(i+1)$. If we define the operator $\mathbb{J}$ via $\mathfrak{C}_2=\mathbb{J}(\mathbb{J}+1)$, then we see that on the tensor product $1^{n-1}\otimes1^{n-1}$, this new operator can be written as
\beq
\label{eq:defJ}
\mathbb{J}=\sum_{i=0}^{n-1}i \proj^{2^{n-1-i}1^{2i}}.
\eeq

\subsection{The \texorpdfstring{$\Rmat$}{R} matrix and the Hamiltonian}
\label{subsec:Rmatrix}

In this subsection, we shall construct the $\Rmat$ matrix of the integrable $\fg$ spin chain with the representation $1^{n-1}$ at each site using the fusion procedure. We let $\para$ be a complex parameter and define the fundamental basic $\Rmat$ matrix that intertwines between two copies of the fundamental representation of $\fg$ to be
\beq
\Rmat(u)\colonequals \frac{u}{\para}+ \per,
\eeq
where $\per$ is the graded permutation operator.
We have the relations
\beq
\label{eq:basicpro}
\Rmat(0)= \per, \qquad \Rmat(\para)=2 \proj^{2^1}, \qquad \Rmat(-\para)=-2 \proj^{1^2},
\eeq
as well as the Yang-Baxter equation
\beq
\label{eq:YBE1}
\Rmat_{12}(u)\Rmat_{13}(u+v)\Rmat_{23}(v)=\Rmat_{23}(v)\Rmat_{13}(u+v)\Rmat_{12}(u).
\eeq
Here, $\proj^{2^1}$ is the projector on the graded symmetric tensor product, while $\proj^{1^2}$ projects on the antisymmetric tensor product. It is possible to generalize \eqref{eq:basicpro} and to write the projector on the $n-1$ fold antisymmetric tensor product of the fundamental representation using the basic $\Rmat$ matrix as
\beq
\label{eq:generalpro}
\proj^{1^{n-1}}=\big(-1\big)^{\frac{(n-1)(n-2)}{2}}\Big(\prod_{k=1}^{n-1}k!\Big)^{-1}\prod_{i<j=1}^{n-1}\Rmat_{ij}\big(\para(i-j)\big).
\eeq
Using the fusion procedure\footnote{See \cite{Zabrodin:1996vm} for a review.}, we can write the non-normalized $\Rmat$ matrix that intertwines between two $1^{n-1}$ representations as
\begin{align}
\label{eq:nonnormalizedRmatrix}
\tilde{\Rmat}^{(n)}(u)\equiv \tilde{\Rmat}^{1^{n-1},1^{n-1}}(u)=&\Big[\prod_{i<j=1}^{n-1}\Rmat_{ij}\big(\para(i-j)\big)\Big]\Big[\prod_{i<j=n}^{2(n-1)}\Rmat_{ij}\big(\para(i-j)\big)\Big]\nonumber\\&\times \Rmat_{1,2n-2}\big(u+(n-2)\para\big)\cdots \Rmat_{1,n}\big(u\big) \nonumber\\&\times \Rmat_{2,2n-2}\big(u+(n-3)\para\big)\cdots \Rmat_{2,n}\big(u-\para\big)\cdots\nonumber\\ &\times \Rmat_{n-1,2n-2}\big(u\big)\cdots \Rmat_{n-1,n}\big(u-(n-2)\para\big)\nonumber\\&\times \Big[\prod_{i<j=1}^{n-1}\Rmat_{ij}\big(\para(i-j)\big)\Big]\Big[\prod_{i<j=n}^{2n-2}\Rmat_{ij}\big(\para(i-j)\big)\Big].
\end{align}
It is straightforward, if tedious, to see using \eqref{eq:YBE1} that the above satisfies the Yang-Baxter equations. Normalizing the expression properly and decomposing it using  the projectors of \eqref{eq:tensordecompositionVV}, we obtain
\beq
\begin{split}
\Rmat^{(n)}(u)&= \frac{1}{(n-1)!\para^{n-1}}\sum_{i=0}^{n-1}\prod_{l=1}^i(u-l\para)\prod_{l=i+1}^{n-1}(u+l\para) \proj^{2^{n-1-i}1^{2i}}\\
&=\frac{\Gamma(n+\frac{u}{\para})}{\Gamma(n)\Gamma(1-\frac{u}{\para})}(-1)^{\mathbb{J}}\frac{\Gamma(1+\mathbb{J}-\frac{u}{\para})}{\Gamma(1+\mathbb{J}+\frac{u}{\para})},
\end{split}
\eeq
where we have used \eqref{eq:defJ}. The last expression\footnote{This $\Rmat$ matrix was first derived in \cite{Kulish:1981gi} for the $\gln{2}$ case.} agrees exactly with formula (1) of \cite{Ferro:2013dga} for $\para=-1$, up to a multiplicative normalization.

At the special value $u=0$, the $\Rmat$ matrix reduces to the graded permutation and thanks to equation \eqref{eq:integrableHamiltonian}, we find that the general integrable Hamiltonian density is 
\beq
\ham= \sum_{i=0}^{n-1}\left[\frac{\alpha}{\para}\Big(\sum_{m=i+1}^{n-1}\frac{1}{m}-\sum_{m=1}^{i}\frac{1}{m}\Big)+\beta\right]\proj^{2^{n-1-i}1^{2i}},
\eeq
for two arbitrary complex number $\alpha$ and $\beta$.
For reasons that shall become apparent in the next section, we wish that $\ham$ should give zero when acting on $2^{n-1}$ and 2 when acting on $2^{n-2}1^2$. The first condition then implies that $\beta=-\frac{\alpha}{\para}\sum_{m=1}^{n-1}\frac{1}{m},$
while the second gives $\alpha=-\para$. Thus, we arrive at the final expression for the Hamiltonian density
\beq
\label{eq:hamiltoniandensityR}
\ham=2\sum_{i=1}^{n-1}h(i)\proj^{2^{n-1-i}1^{2i}},
\eeq
where $h(i)\colonequals \sum_{m=1}^i\frac{1}{m}$ are the harmonic numbers. This Hamiltonian density is a finite dimensional case of the one loop dilatation generator appearing in $\mathcal{N}=4$ Super Yang-Mills \cite{Beisert:2004ry}.

\subsection{The local length-changing operator}
\label{subsec:locallengthchanging}

We are now looking for an operator $\ya:V\rightarrow V\otimes V$ such that the associated Hamiltonian density \eqref{eq:localhamiltoniandensity} agrees with the one \eqref{eq:hamiltoniandensityR} derived from the $\Rmat$ matrix. In order to preserve the $\fg$ symmetry, we require that $\ya$ commutes with the Cartan subalgebra $\fh\subset \fg$. For $\slmn{n}{1}$, the charge of the vector $e_{n+1}$ is equal to the sum of the charges of all the bosonic vectors $(e_i)_{i=1}^n$. Therefore, when going from one to two sites, we need always to remove one $e_{n+1}$ and add all the others in some appropriate order. In the hole basis \eqref{eq:definitionnewbasis}, this means that $\ya$ takes the vector $\ket{k_1,\ldots, k_r}$ and maps it to a linear combination of $\ket{k_{\sigma(1)},\ldots, k_{\sigma(m)}}\otimes \ket{k_{\sigma(m+1)},\ldots, k_{\sigma(r)}}$, where $\sigma$ is some arbitrary permutation and $m$ can run from $1$ to $r-1$. Essentially, it just distributes the holes on two sites. In particular, this implies that $\ya \ket{k}=0$. We can now try to find the action of $\ya$ on the highest weight state that has two holes. Invariance under $\fh$ implies
\beq
\ya \ket{n-1,n}=\gamma_1\ket{n-1}\otimes \ket{n}+\gamma_2\ket{n}\otimes \ket{n-1}
\eeq
for arbitrary complex $\gamma_i$. We shall at this point impose two additional requirements that are motivated from the study of the $n=2$ system, namely that $\ya$ commutes with all bosonic generators of $\fg$ and that its anticommutator with the simple negative fermionic root operator is given by 
\beq
\label{eq:nontrivialanticommutator}
\Big(\ya E_{n+1,n}+\Delta(E_{n+1,n})\ya\Big)v =v\otimes \ket{n}-\ket{n}\otimes v, \qquad \forall v \in V .
\eeq
The first requirement sets $\gamma_2=-\gamma_1$ and we normalize $\gamma_1$ to one. By using the condition \eqref{eq:nontrivialanticommutator} and by requiring that $\ya$ commutes with the other simple generators, we guess the following expression\footnote{While \eqref{eq:localactionofq} seems superficially similar to the expression \eqref{eq:highestweightvectors} for the highest weight vectors, observe that the right-hand side of \eqref{eq:localactionofq} is always graded antisymmetric, while the $v_i$ are symmetric for $i$ even and antisymmetric for $i$ odd.} for the action of $\ya$:
\beq
\label{eq:localactionofq}
\ya\ket{k_1,\ldots,k_r}=\sum_{m=1}^{r-1}(-1)^{rm}\sum_{\sigma\in S_{m,r-m}}(-1)^{|\sigma|}\ket{k_{\sigma(1)},\ldots,k_{\sigma(m)}}\otimes \ket{k_{\sigma(m+1)},\ldots,k_{\sigma(r)}}.
\eeq
We remind that due to \eqref{eq:defquotientS}, the $k_{\sigma(i)}$ are all arranged as in \eqref{eq:definitionnewbasis}, i.e. such that $k_{\sigma(i)}< k_{\sigma(i+1)}$ for $i=1,\ldots m-1$ and for $i=m+1,\ldots ,r-1$.  

It is trivial to check, using \eqref{eq:actionofsymmetryholebasis}, that $\ya$ commutes with all bosonic generators. Furthermore, an explicit computation confirms that \eqref{eq:nontrivialanticommutator} does indeed hold and that furthermore $\ya$ anticommutes with the fermionic raising generators. Thus, we can safely say that $\Ya$ commutes in the graded sense with the full action of $\fg$. Furthermore, we have checked that $\ya$ obeys equation \eqref{eq:localnilpotency} with $\chi=0$, thus guaranteeing that the supercharge is nilpotent.

We now wish to define the local length-lowering operator by considering the adjoint of $\ya$. For this purpose, we introduce the scalar product on $V_F$ by $\form{e_i}{e_j}=\delta_{ij}$. This together with \eqref{eq:definitinoadjoint} implies that $E_{ij}^{\dagger}=(-1)^{|i|(|j|+1)}E_{ji}$. In order for this to be compatible with the coproduct, we require that the scalar product is generalized to tensor products of $V_F$ as
\beq
\form{e_{i_1}\otimes \cdots \otimes e_{i_r}}{e_{j_1}\otimes \cdots \otimes e_{j_r}}=(-1)^{\sum_{k=0}^{r-1}|i_{r-k}|\sum_{l=1}^{r-1-k}|j_l|}\delta_{i_1j_1}\cdots \delta_{i_rj_r}.
\eeq
For the orthogonal basis \eqref{eq:definitionnewbasis} of $V$, we get the following expressions for the square of the norm:
\beq
\label{eq:scalarproductk}
\form{\ket{k_1,\ldots,k_r}}{\ket{k_1,\ldots,k_r}}=(-1)^{\frac{(r-2)(r-1)}{2}}\frac{(r-1)!}{(n-1)!}.
\eeq
It follows after a direct computation using \eqref{eq:definitinoadjoint} that
\begin{multline}
\ya^{\dagger}\ket{k_1,\ldots,k_m}\otimes \ket{k_{m+1},\ldots,k_r}=\\=(-1)^{rm}\frac{(m-1)!(r-m-1)!}{(r-1)!(n-1)!}(-1)^{|\sigma|}\ket{k_{\sigma(1)},\ldots,k_{\sigma(r)}}.
\end{multline}
where $\sigma$ is a permutation that brings the $k_j$ in the correct order.

We now have all the ingredients to write the Hamiltonian density for the length-changing operators. Since $\ya$ is fermionic for our grading, we use \eqref{eq:localhamiltoniandensity} with $\omega=1$, normalized to:
\beq
\label{eq:normalizedhamiltonian}
\ham(\ya)=(n-1)!\Big[\big(\yas\otimes \mathbbm{1}\big)(\gr\otimes \ya)+(\gr\otimes \yas)\big(\ya\otimes \mathbbm{1}\big)+\ya\yas+\frac{1}{2}\big(\yas\ya\otimes \mathbbm{1}+\mathbbm{1}\otimes \yas\ya\big)\Big].
\eeq
In order to compare with \eqref{eq:hamiltoniandensityR}, we need to compute the action of $\ham(\ya)$ on the highest weight vectors $v_i$ of \eqref{eq:highestweightvectors}. The case of $v_0$ is simple since $\ya \ket{k}=0$ and $\yas \ket{n}\otimes \ket{n}=0$, so that
\beq
\ham(\ya)v_0=0.
\eeq
For the highest weight vector $v_1$ we find
\beq
\ham(\ya)v_1=\ya \yas v_1=2\ya \ket{n-1,n}=2v_1.
\eeq
For $v_2$, we first see that $\yas v_2=0$ and after some algebra arrive at  $\ham(\ya)v_2=3v_2$ as required.
The action on $v_3$ is already quite complicated, but a careful computation gives an eigenvalue of $\frac{11}{3}=2h(3)$ in full agreement with \eqref{eq:hamiltoniandensityR}. With the help of a computer algebra program, we were able to check that \eqref{eq:normalizedhamiltonian} agrees with \eqref{eq:hamiltoniandensityR} when acting on the highest weight vectors $v_i$ with $i\leq 8$, so that we are confident that the relation holds in general.

\section{Conclusion}
\label{sec:conclusion}

In this article we examined the Bethe equations of the rational and trigonometric $\glnm{n}{m}$ spin chains for degeneracies and could classify the cases for which dynamic, i.e. length-changing, supersymmetry in the sense of section \ref{sec:dynsusy} could be expected to be present. We then took the simplest new example for which the Bethe equations hinted at the existence of such a supersymmetry, namely the $\slmn{n}{1}$ spin chains made out of copies of the $(n-1)$-fold antisymmetric tensor product of the fundamental, and constructed the supercharges explicitly, checking in the process that they indeed reproduce the integrable Hamiltonian. It would be interesting to find a general way of constructing the supercharges that is valid for all candidate representations.

The procedure of analyzing the Bethe equations and constructing the supercharges can be generalized in several ways. The most promising one would be to allow for different representations to be present at each site of the chain and to introduce inhomogeneities in the transfer matrix. Of course, for the concept of dynamic supersymmetry to still make sense, some notion of cyclicity has to remain, which means that the most promising approach is to restrict oneself to alternating spin chains. For those cases, the generalization of section \ref{sec:bethe} is indeed straightforward but so far, we have not been able to write the supercharges explicitly for any new model\footnote{See \cite{Zwiebel:2009vb} for an example appearing in the context of the ABJM duality.}. Often, the main obstacle is related to the fact that the cyclicity condition gets replaced by a more stringent condition which requires that the sectors of interest be the combined eigenspaces of eigenvalue one of the transfer matrices\footnote{We recall that for alternating spin chains one has to construct two transfer matrices, depending on the kind of representation that is inserted in the auxiliary space and that the true transfer matrix which reduces to the shift operator is given by the product of the two.} $\tra_L$ and $\bar{\tra}_L$ at some special values of the spectral parameter.  Unfortunately, these operators at those special values of the spectral parameter are often non-unitary and non-diagonalizable, which makes the precise description of the supercharges quite complicated. 

Another important avenue of future research is the precise nature of the relationship between integrability and dynamic supersymmetry. We expect from the analysis of the Bethe equations that the transfer matrix commutes with the supercharges and, for the specific examples for which the supercharges have been constructed, this has been checked at level of the operators for small spin chain lengths. So far, there is no proof that this should hold for all lengths and it is to be hoped that elucidating this issue would also give one a way of finding a set of conditions for the local supercharges $\ya$ so that the resulting Hamiltonian is integrable. 

Finally, gauge theory suggests that we should also generalize to spin chains that are long-ranged and for which the Hamiltonian itself is length-changing.  For this purpose the works of \cite{Zwiebel:2008gr} and \cite{Bargheer:2008jt,Bargheer:2009xy} might provide some clue. Of course, if might also be possible to find a non-perturbative way to freeze the length-changing via some complicated basis transformation generalizing \cite{Fendley:2002sg, Fendley:2003je, Hagendorf:2012fz}.
We hope to be able to answer at least some of these questions in future works.

\section*{Acknowledgments}
The authors thank Rouven Frassek, Sergey Frolov, Nils Kanning, Yumi Ko, Matthias Staudacher, Zengo Tsuboi, Matthias Wilhelm and especially Christian Hagendorf for inspiring discussions and insightful suggestions. Furthermore VM is  grateful to the Kavli IPMU, Universit\'e Catholique de Louvain and Sogang University for the kind hospitality while working on parts of the manuscript



\providecommand{\href}[2]{#2}\begingroup\raggedright\endgroup

\end{document}